\documentclass[10pt,letterpaper]{article}
\usepackage{opex3}
\usepackage{color}
\usepackage{threeparttable}
\usepackage{amsthm}
\usepackage{amsmath}
\usepackage{latexsym}
\usepackage{amsfonts}
\usepackage{amssymb}
\usepackage{bbm,dsfont}
\usepackage{graphicx}
\usepackage{subfigure}
\usepackage{tikz}
\usepackage{placeins}
\usepackage{epstopdf}
\usepackage[square, comma, sort&compress, numbers]{natbib}

\newcommand{\ket}[1]{|#1\rangle} 
\newcommand{\kb}[2]{|#1\rangle\langle#2|} 
\newcommand{\tr}[2]{\text{tr}_{#2}\left\{#1\right\}}
\newcommand{\id}{\mathbbm{1}} 
\newcommand{\meas}{{\rm d\,}}
\newcommand{\U}{\mathcal{U}}

\begin{document}

\title{Non-Markovian discrete qubit dynamics}

\author{Jun Sun$^{1,2}$, Yong-Nan Sun$^{1,2}$, Chuan-Feng Li$^{1,2,7}$, Guang-Can Guo$^{1,2}$, Kimmo Luoma$^{3,5}$ and Jyrki Piilo$^{4,6}$}

\address{$^1$Key Laboratory of Quantum Information, University of Science and Technology of China CAS Hefei, 230026, China\\
$^2$Synergetic Innovation Center of Quantum Information and Quantum Physics, University of Science and Technology of China, Hefei, 230026, China\\
$^3$Institut f{\"u}r Theoretische Physik, Technische Universit{\"a}t Dresden,
D-01062, Dresden, Germany\\
$^4$Turku Centre for Quantum Physics, Department of Physics and
Astronomy, University of Turku, FI-20014, Turun Yliopisto, Finland\\
$^5$kimmo.luoma@tu-dresden.de\\
$^6$jyrki.piilo@utu.fi\\
$^7$cfli@ustc.edu.cn

}


\begin{abstract}
The study of open quantum systems is important for fundamental issues of quantum physics
 as well as for technological applications such as quantum information processing.
 The interaction of a quantum system with it's environment
 is usually detrimental for the quantum properties of the system
 and leads to decoherence. However, sometimes a coherent partial exchange of
 information takes place between the system and the environment and the dynamics of
 the open system becomes non-Markovian. In this article we study discrete open quantum
 system dynamics where single evolution step consist of local unitary transformation
 on the open system followed by a coupling unitary between the system and the environment.
 We implement experimentally a local control protocol for controlling the transition from Markovian
 to non-Markovian dynamics.
\end{abstract}




\section{Introduction}
Whenever quantum system is not perfectly
isolated (up to experimental accuracy) it has to be treated as an open system~\cite{breuer_theory_2007}.
The description of the open system dynamics is based on a family of completely positive and trace preserving maps (CPT),
so called dynamical maps, governing
the evolution of the state of the open system as $\rho \mapsto \rho'=\Phi\rho$~\cite{alicki_quantum_1987}.
Dynamical maps are a standard tool to describe dechorence and dissipation phenomena.
If the family of dynamical maps has semi-group property then its properties and structure
are well known. The generator of the quantum dynamical semigroup in time continuous case is
the famous Gorini-Kossakowski-Sudarshan-Lindbald (GKSL) master equation~\cite{gorini_completely_1976,lindblad_generators_1976}
which is the workhorse
of the open quantum systems research~\cite{haken_quantum_1999}. Open systems falling into a category where their dynamics
can be described with quantum dynamical semigroup are called Markovian~\cite{breuer_non-markovian_2015}.

In general the dynamical map is less structured, meaning
the dynamical map may not necessarily be divisible, not even with positive maps~\cite{wolf_dividing_2008}.
Recently due to efforts to quantify quantum non-Markovianity \cite{wolf_assessing_2008,breuer_measure_2009,
rivas_entanglement_2010,luo_quantifying_2012,lorenzo_geometrical_2013,chruscinski_degree_2014,bylicka_2014,pollock_complete_2015}
and technological advances \cite{liu_experimental_2011,Smirne_2011,Chiuri_2012,Bernardes_2015},
non-Markovian quantum processes have become a central topic in the study of open
quantum systems.
As a matter of fact, there exists different notions of quantum non-Markovianity where the key concepts include the breakdown of divisibility of the map. In general, reliastic quantum systems interact and exchange information with the environment. The engineering of the decoherence and flow of information between an open
quantum system and its environment is the key to control the transition from Markovian
dynamics to a regime with quantum memory effects

It is well known that by using dynamical decoupling techniques, where fast pulses
are applied to the open system locally, the environmental effects may be eliminated \cite{viola_dynamical_1999,
uhrig_keeping_2007,lidar_review_2014}. This is
one form of quantum control where in general controls are applied to
the system of interest in order to minimize or maximize some objective functional \cite{jirari_quantum_2006,
rebentrost_optimal_2009,lapert_singular_2010,schmidt_optimal_2011,hwang_optimal_2012,tai_optimal_2014}.
Usually, the transition from Markovian to non-Markovian dynamics is related to strong coupling
between the open system and the environment or
complex spectra of the environment.
In this work we report on experimental results regarding to the transition from
Markovian to non-Markovian dynamics using local controls on open system coupled to
a Markovian structureless environment. Our approach bears resemblance to the dynamical decoupling since we assume that the local control
acts instantaniously but on the other hand we do not apply the control operations in rapid succession.
Our goal
is to transform the free evolution between local controls to be non-Markovian.

\section{Theory}
We conduct our experiments with single photons. The
polarization states of the photon $\ket{H},\,\ket{V}$ form a qubit and
the frequency degree of freedom acts as an environment for the qubit.
We prepare the initial state of the photon to be either
$\ket{H,\omega}$ or $\ket{V,\omega}$.
The frequency distribution of the photon
$\chi(\omega,\omega)\equiv\vert{\chi(\omega)}\vert^2$ is well described with a
single Gaussian with standard deviation $\sigma$.
The spectral distribution is
centered around the laser wavelength $\lambda_0$.
Local control operations are unitaries acting only on the polarization degrees
of freedom, they are implemented by half-wave plates described by the following unitary
\begin{align}\label{eq:1}
  C_{\eta} =& \eta^\frac{1}{2}\sigma_z+(1-\eta)^\frac{1}{2}\sigma_x,
\end{align}
where $\eta$ is controlled by the tilt angle and Pauli matrices $\sigma_i$
are written in the polarization basis. For each $\eta$ we choose, the corresponding
angle of the half-wave plate is given by $\varphi_\eta=1/2\arccos{(\eta^\frac{1}{2})}$.
Simple theoretical model for the quartz plate is given by the following
unitary coupling the qubit and the environment Hilbert spaces
\begin{align}\label{eq:2}
  U_{\delta t}=&\sum_{k=H,V}\int\meas \omega exp(i n_k\omega \delta t)\kb{k}{k}\otimes\kb{\omega}{\omega},
\end{align}
where $\delta t$ is the interaction time of the photon with the environment and  $\Delta n =n_H-n_V= 0.008995$
is the difference of the polarization dependent indices of refraction. Interaction
time $\delta t$ is related to the quartz plate thickness by $L = c\delta t$, where
$c$ is the speed of light. We express the interaction time of each of the
quartz plates in terms of effective path difference $\Delta L=\Delta n L$ in
the units of $\lambda_0$. Total unitary, or {\it operating unit}, corresponding to a single step is now described as $U_i \equiv U_{\delta t_i}(C_{\eta_i}\otimes\id_E)$.
The dynamical map can be written as
\begin{align}\label{eq:3}
  \Phi_n(\rho_0)=&\tr{\U(n)(\rho_0\otimes\chi) \U^\dagger(n)}{E},
\end{align}
where $\U(n) = \prod_{i=1}^nU_{\delta t_i}(C_{\eta_i}\otimes\id_E)$.
Non-Markovianity is detected by the temporal increase of distinguishability between the quantum states.
Distinguishability between two quantum states $\rho_1,\rho_2$ is defined as
$D(\rho_1,\rho_2) = \frac{1}{2} \vert\vert \rho_1-\rho_2 \vert\vert$
where $\vert\vert\cdot\vert\vert_1$ is the trace norm. Distinguishability is a monotonically
decreasing function under positively divisible quantum dynamical maps \cite{breuer_measure_2009}. We take this to be
our definition of non-Markovianity. If we define the increment of the
trace distance evolution as $\Delta_{1,2}(n)=D(\rho_1(n),\rho_2(n))-D(\rho_1(n-1),\rho_2(n-1))$, where
$\rho_i(n)$ is $\Phi_n(\rho_i(0))$, we obtain a measure for non-Markovianity as
$\mathcal{N}(n)=\max_{\rho_{1,2}(0)}\sum_{\Delta_{1,2}(n)>0}\Delta_{1,2}(n)$. A lower bound for the
measure is obtained just with a single pair that shows non-monotonic distinguishability
evolution. In this experiment the initial state pairs were taken to be
$\rho_1(0) = \kb{H}{H}$ and $\rho_2(0)=\kb{V}{V}$.
We know from earlier studies that when frequency distribution is a single
Gaussian it will lead to Markovian dynamics without local control~\cite{liu_experimental_2011,luoma_discrete_2015}.
The above theoretical model includes the local control of the open system
and also the interplay of the local controls and the correlations between the open
system and the environment. Specifically we can explain how by adding suitable local
control we can control the transition from Markovian to non-Markovian dynamics
even for ``Markovian'' environment.
This should be contrasted with a phenomenological
approach where one adds the local control Hamiltonian to the the GKSL master equation.
Such a master equation can not clearly be responsible
for the increase of the trace distance.

\section{Experiment}

\begin{figure}[htbp]
\centering\includegraphics[width=9cm]{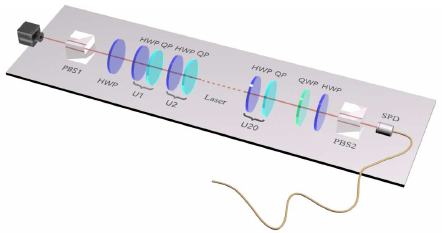}
\caption{\label{fig:ExpSetup} The experimental setup. Key to the components:
  HWP--half-wave plate, QP--quartz plate, QWP--quarter-wave plate,
  PBS--polarizing beam splitter and SPD--single photon detector. The laser was
  attenuated into single photon states and initialised by PBS1 and the first half-wave
  plate into $\kb{H}{H}$ and $\kb{V}{V}$. After 20 steps of local control and
  interactions with the environment implemented by half-wave plates and quartz plates,
  the states are finally analysed by a quarter-wave plate,
  a half-wave plate and PBS2. }
\end{figure}

The experimental setup is shown in Fig.~\ref{fig:ExpSetup}. A femtosecond pulse
(with a duration of about
150 fs, the operation wavelength at $\lambda_0 = 800$ nm and with a repetition rate of about 76 MHz)
generated from a Ti sapphire laser is used to generate arbitrary pure qubit states. The
full width at half maximum (FWHM) of the laser is 6 nm and the laser was attenuated in order to
neglect the two-photon coincidences. The photon states are initialized by PBS1 and the first
half-wave plate into $\kb{H}{H}$ and $\kb{V}{V}$. We also use PBS2 together with the last half-wave plate
and the quarter-wave plate for tomography.
The environment is characterized by a single
Gaussian peak with standard deviation $\sigma = 2.55$ nm. We use a half-wave plate and a quartz plate to realize the unitary control
and decoherence which, when combined,  is called a single operating unit $U_1$.
In a single experiment, we have
20 sets of units from $U_1$ to $U_{20}$ with the same angle of the HWPs (with respect to
the optical axis)
and the same thickness
of the quartz plates. We can control the dephasing strength with the
thickness of the quartz plate and have two different thichnesses at our disposal: $7.111$ mm and
$10.667$ mm. We can choose which local control we implement by choosing the angle
of the HWPs. We use five different angles $\varphi_\eta$:
$9^\circ,\, 18^\circ,\, 22.5^\circ,\, 28^\circ$, and $33^\circ$, corresponding to
five different values of $\eta$:
$0.9045$, $0.6545$, $0.5000$, $0.3127$ and $0.1654$, respectively.

\begin{figure}[htbp]
\centering\includegraphics[width=7cm]{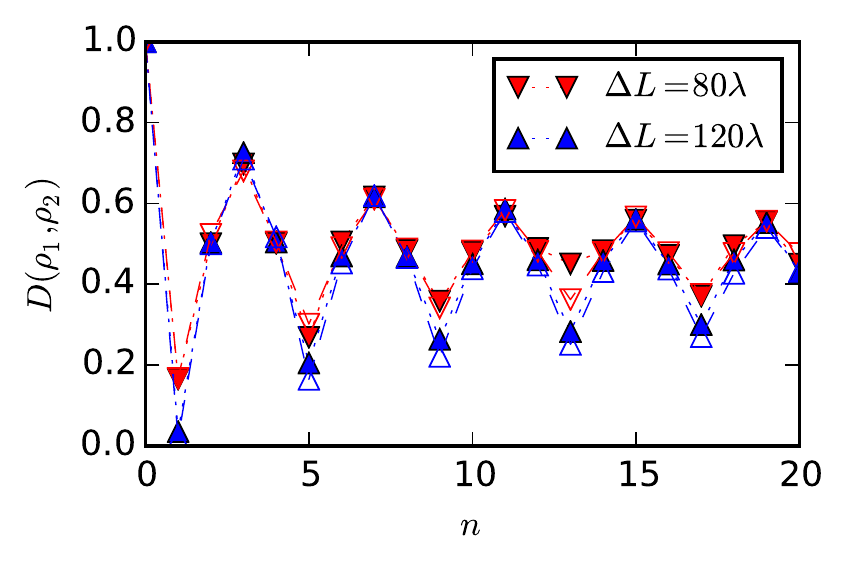}
\caption{\label{fig:TD_varL} Trace distance dynamics for two different quartz plate thicknesses $\Delta L$ for
  parameters $\Delta n = 0.008995$, FWHM = $6$ nm, $\lambda = 800$ nm and $\eta=0.5$. Initial state pair
  is $\ket{H}$, $\ket{V}$.
  Filled markers are for experimental results and empty markers are for numerical simulations. }
\end{figure}

\begin{figure}[htbp]
\centering\includegraphics[width=7cm]{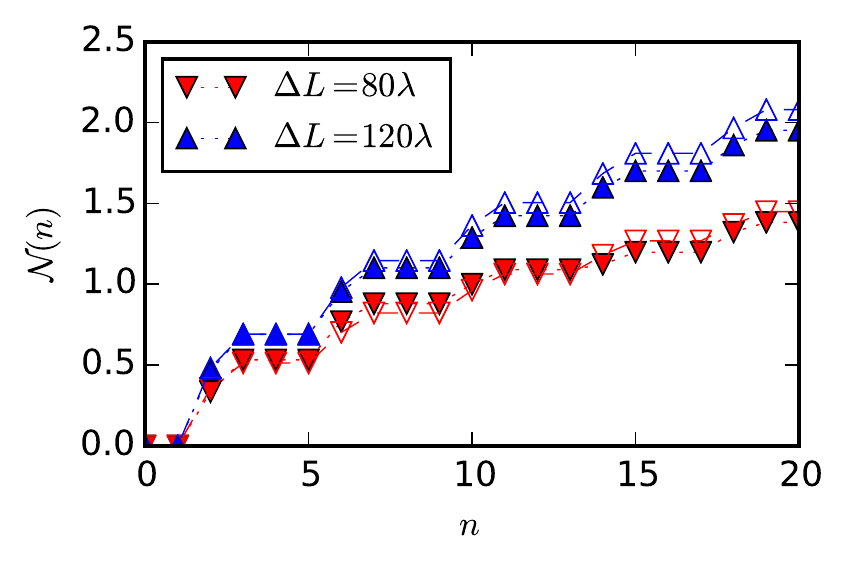}
 \caption{\label{fig:NM_varL} Non-Markovianity for two different quartz plate thicknesses $\Delta L$ for
  parameters $\Delta n = 0.008995$, FWHM = $6$ nm, $\lambda = 800$ nm and $\eta=0.5$. Initial state pair
  is $\ket{H}$, $\ket{V}$.
  Filled markers are for experimental results and empty markers are from numerical simulations.}
\end{figure}
\begin{figure}[htbp]
 \centering\includegraphics[width=7cm]{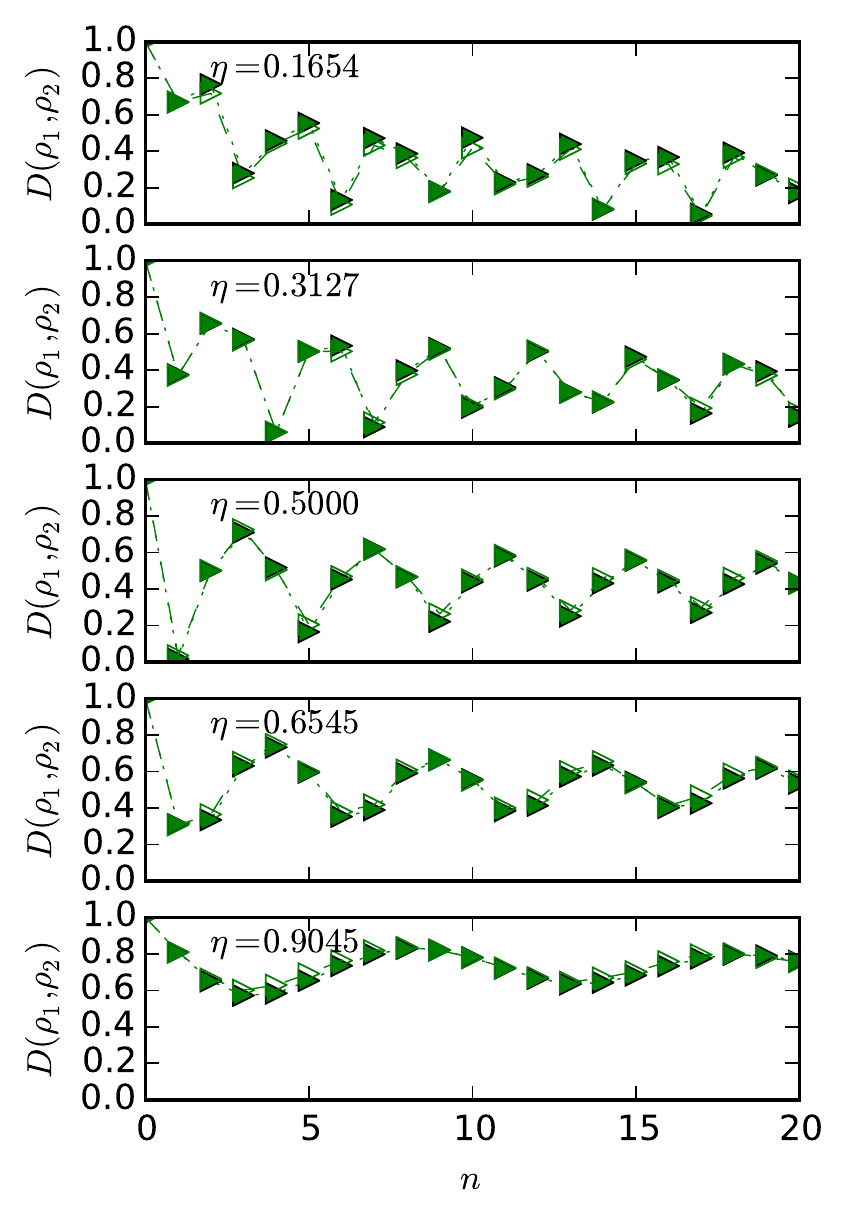}
  \caption{\label{fig:TD_VAR_ETA} Trace distance dynamics for different values of $\eta$ for
  parameters $\Delta n = 0.008995$, FWHM = $6$ nm, $\lambda = 800$ nm and $\Delta L = 120\lambda$. Initial state pair
  is $\ket{H}$, $\ket{V}$. Filled markers are for the experimental data
  and empty markers are for the theoretical calculation.}
\end{figure}

\begin{figure}[htbp]
\centering\includegraphics[width=7cm]{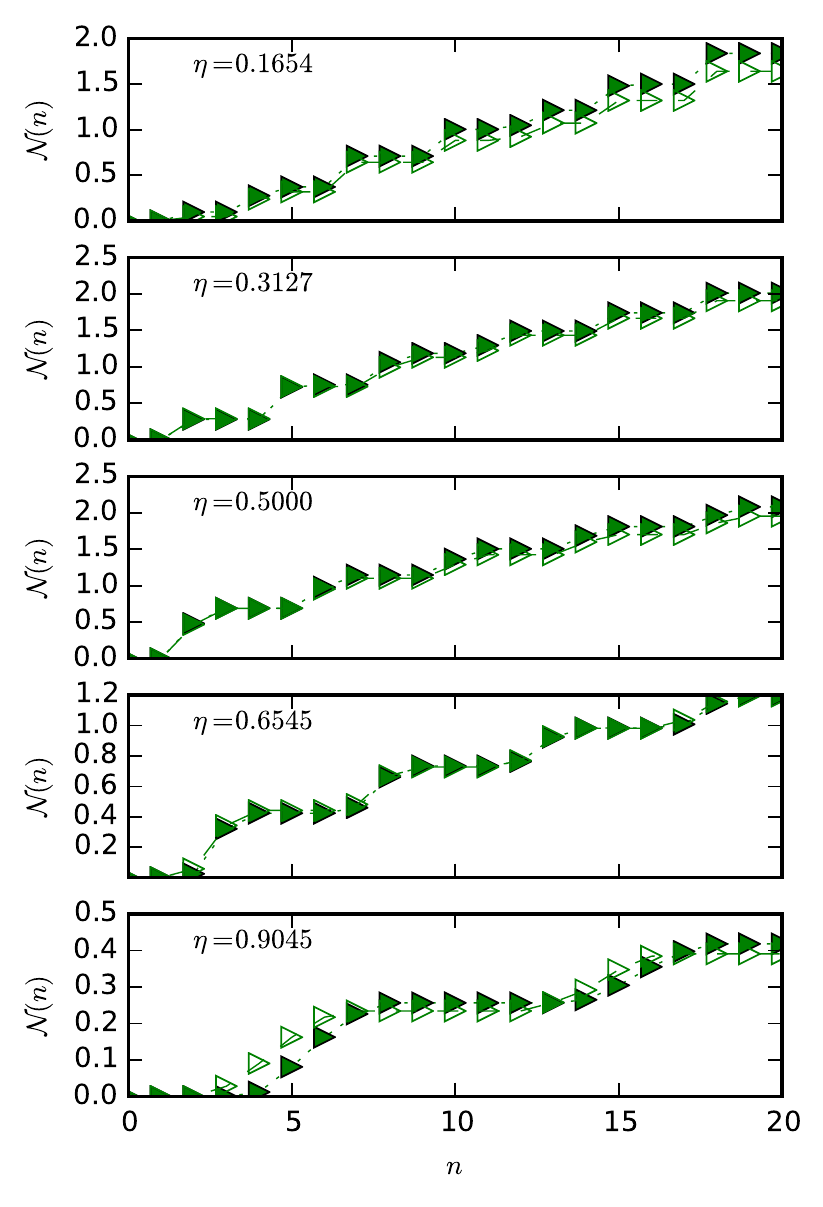}
  \caption{\label{fig:NM_VAR_ETA} Non-Markovianity for different values of $\eta$ for
  parameters $\Delta n = 0.008995$, FWHM = $6$ nm, $\lambda = 800$ nm and $\Delta L = 120\lambda$. Initial state pair
  is $\ket{H}$, $\ket{V}$. Filled markers are for the experimental data
  and empty markers are for the theoretical calculation.}
\end{figure}

\section{Result}
The effect of the pure dephasing unitary of Eq.~\eqref{eq:2} without local control onto the open
system is to shrink the Bloch sphere
towards the z-axis along the directions parallel to $x,y$-plane, given our choice of
initial environment state. Whenever $0\leq \eta < 1$ the local control $C_\eta$ is not diagonal
in the $H,V$-basis and thus the effect of the local control is non-trivial.  We also note that
if $\eta=0$  spin echo occurs in the open system dynamics, ie. a perfect recovery of the initial state
after of even number steps.

First results show what is the effect of the interaction time for
fixed local control ($\eta_i=\eta=\frac{1}{2}$) and for equally thick
quartz plates at every step. In Figs. \ref{fig:TD_varL}, \ref{fig:NM_varL}
we have plotted the dynamics of the trace distance for initial qubit state pair
$\ket{H}$ and $\ket{V}$ and the associated lower bound on the
non-Markovianity.  One can see from Fig. \ref{fig:TD_varL} how the longer interaction
time is causing more dephasing during the initial step. This is easy to understand because
for the first step the effect of the local control is just to rotate the initial states and the
dephasing dynamics then moves both initial states towards the z-axis, hence making them
less distinguishable. One should also note that the $\eta=\frac{1}{2}$ rotates both initial states
on the equator of the Bloch sphere and since the states are pure the distance parallel to the
$x,y$-plane from the surface of the Bloch sphere to the z-axis is maximal.
After the second step the local control starts to have an effect on the correlation build-up
between the system and the environment. From Fig.~\ref{fig:TD_varL}  we can see
that the more we dephase on the first step, the stronger the non-Markovianity is,
Fig.~\ref{fig:NM_varL}. Interestingly the values of maximums of the trace distance values coincide
but the minimas do not for $\Delta L = 80\lambda$ and $\Delta L = 120\lambda$. This means
that for these cases the local control
sets an upper bound on how well distinguishability can be restored. On the other hand, the dephasing strength
sets a lower bound how indistinguishable states can become under the dynamics.

Next we study the effects of the $\eta$ while keeping other parameters fixed but still keeping
the quartz plate thickness and $\eta$ uniform trough the whole experiment.
We have plotted the trace distance dynamics on Fig.~\ref{fig:TD_VAR_ETA} and the associated
non-Markovianity measure on Fig.~\ref{fig:NM_VAR_ETA}. We have kept the initial state
pair as before. Whenever $\eta\neq\frac{1}{2}$ the distance parallel to
$x,y$-plane from the surface of the Bloch sphere to the
z-axis is not maximal after the first local control operation. This can be seen from Fig.
\ref{fig:TD_VAR_ETA} where the decrease of the trace distance after the first step
is maximal for $\eta=\frac{1}{2}$ and for the used initial state pair.
When $\eta=1$ theoretical model predicts Markovian dynamics. The behavior
of the trace distance is in accordance with this finding since the structure
becomes less oscillating as $\eta$ is increasing. For $\eta < 0.9045$ the
behavior of trace distance dynamics is periodic with a period of four. This is
in accordance with our experimental results. Also, when $\eta=0$ then
the trace distance dynamics should have a period of two. We can see that
for $\eta=0.1654$ the periodicity is not so clear anymore, for example
the first two minimas are two steps apart ($n=1,\,n=3$). The dynamics in all
of the cases is non-Markovian as can be seen from Fig.~\ref{fig:TD_VAR_ETA}.
When the units $U_i$ are chosen uniformly ie. $U_i=U$, we verified numerically that in the
large step number limit the open system dynamcis approaches a steady state
logarithmically. The steady state is given by a statistical mixture that is diagonal
in the eigenbasis of the local control $C_i=C$.

\section{Conclusions}
We show that using local control only on the open system it is possible to control
the transition from Markovian to non-Markovian dynamics. From the point of view of
quantum control our work shows that it is necessary to consider the control problems in
terms of the full open system and environment picture since the controlled dynamics
might not be Markovian even for Markovian environment.
In the sight of open quantum system's research we show that the influence of
the system-environment correlations can be controlled locally leading also to the control of amount of memory effects.
Indeed, by locally rotating the quantum state we can engineer non-Markovian quantum evolution.

\section*{Acknowledgments}
This work was supported by the Magnus Ehrnrooth Foundation
and Academy of Finland (Project no. 287750), National Natural Science Foundation of China (Nos. 61327901, 11274289, 11325419),
the Strategic Priority Research Program (B) of the Chinese Academy of Sciences (Grant No. XDB01030300).
C.F.L. and J.P. acknowledge financial support from the EU Collaborative project QuProCS (Grant Agreement 641277).

\end{document}